\documentclass{mn2e}
\usepackage{times}
\usepackage{epsf}

\newcommand{\cl}{\hbox{CL\,0016$+$16}}
\newcommand{\asca}{\textit{ASCA}}
\newcommand{\chandra}{\textit{Chandra}}
\newcommand{\rosat}{\textit{ROSAT}}
\newcommand{\xmm}{\textit{XMM-Newton}}

\begin{document}

\title[\cl\ measured with \xmm]{The temperature and
 distribution of gas in \cl\ measured with \xmm}

\author[D.M. Worrall \& M. Birkinshaw]{D.M. Worrall and M. Birkinshaw\\
Department of Physics, University of Bristol, Tyndall Avenue,
Bristol BS8~1TL}

\maketitle

\label{firstpage}

\begin{abstract}
We present results of a 37~ks observation of \cl\ with the \xmm\ EPIC
instrument. Within 1.5 arcmin of the cluster centre we measure a gas
temperature of $kT = 9.13^{+0.24}_{-0.22}$~keV and an abundance of
$0.22^{+0.04}_{-0.03}$ times the solar value ($1\sigma$
uncertainties).  This significant improvement over previous
measurements has allowed us to revise the estimate of the Hubble
constant based on \cl\ to $68 \pm 8 \ \rm km \, s^{-1} \, Mpc^{-1}$
(random error only), close to the value from the Hubble Space
Telescope distance-scale project.  The total gravitating mass within a
radius of 248~kpc of the cluster centre is in good agreement with that
found from gravitational lensing over the same region, supporting the
assumption of isothermal gas in hydrostatic equilibrium. The gas mass
fraction of $0.13 \pm 0.02$ is in remarkable agreement with that given
by cosmological parameters for the Universe as a whole, suggesting
that \cl\ is a fair sample of the matter content of the Universe.
While there is no spectral or spatial evidence to suggest a cooling
flow in \cl, we find an asymmetrical central X-ray structure which may
have a harder spectrum than the cluster as a whole, and be evidence
for some merger activity, in addition to a previously reported
asymmetry to the west of the cluster.  The nearest companion cluster
to \cl\ is measured to have a gas temperature of $kT =
3.8^{+0.3}_{-0.3}$~keV and an abundance of $0.6^{+0.3}_{-0.2}$ times
the solar value ($1\sigma$ uncertainties).  We also present spectral
data for the companion quasar whose spectrum was confused with \cl\ in
previous \asca\ data.
\end{abstract}


\section{Introduction}
\label{sec:intro}

The galaxy content of cluster \cl\ was described by Koo (1981)
following its discovery by Kron (1980) in a photometric survey for
faint galaxies. Its high redshift, $z=0.5455$ (Dressler \& Gunn 1992),
evident richness, high X-ray luminosity ($L_{\rm X,~0.5-4.5~keV}
\approx 2 \times 10^{45}$ ergs s$^{-1}$; White, Silk \& Henry 1981),
and large Sunyaev-Zel'dovich effect (Birkinshaw, Gull \& Moffet 1981)
led to extensive work on its properties as the prototypical example of
a distant rich cluster.

More recent X-ray and optical observations of the field of \cl\ have
found other nearby clusters of galaxies of lower mass but similar
redshift (Hughes, Birkinshaw \& Huchra 1995; Connolly et al. 1996;
Hughes \& Birkinshaw 1998a), suggesting that \cl\ is the dominant
member of a supercluster at $z \sim 0.5$. A quasar a few arcmin north
of \cl\ (Margon, Downes \& Spinrad 1983) is also associated with this
structure.

The gas properties of \cl\ have been the subject of particular
study. Two reasons for this stand out: the intracluster gas can be
used to study the cluster mass and baryonic mass fraction, and the
combination of X-ray and Sunyaev-Zel'dovich data can be used to
estimate the cluster's distance and hence the Hubble constant or other
cosmological parameters.

Grego et al. (2001) used X-ray data and the Sunyaev-Zel'dovich effect
of \cl\ to measure a gas mass fraction, $f_{\rm g}$, that is close to
the value seen in nearby clusters, implying that there is little
evolution in $f_{\rm g}$ between $z \sim 0.5$ and today. If cluster
gas is the dominant baryonic component of \cl, and \cl\ is a fair
sample of the matter content of the Universe, $f_{\rm g}$ will be
close to the baryonic mass fraction of the Universe, $\Omega_{\rm
b}/\Omega_{\rm m}$. However, the temperature of the cluster gas is a
necessary component of the calculation of both the gas and total
masses, and improved measurements of the gas temperature would be
expected to improve the reliability of cosmological conclusions drawn
from studies of the mass components of the cluster.

Good knowledge of the cluster temperature would also allow a
comparison between the mass derived from the assumption that the
cluster gas is in hydrostatic equilibrium and the mass measured by
Smail et al. (1997) using a gravitational shear technique. This could
set limits on the three-dimensional shape of \cl, which is a source of
systematic error in the use of any cluster as a cosmological tracer.

Three independent Sunyaev-Zel'dovich effect measurements of \cl\
(Hughes \& Birkinshaw 1998b; Reese et al. 2000; Grainge et al. 2002)
have been combined with X-ray data to measure the distance of the
cluster. All three lead to values for the Hubble constant of about $47
\ \rm km \, s^{-1} \, Mpc^{-1}$ assuming $q_0 = {1 \over 2}$ (this
would be $57 \ \rm km \, s^{-1} \, Mpc^{-1}$ for a cosmology with
$\Omega_{\rm m} = 0.3$, $\Omega_\Lambda = 0.7$). The low value of the
Hubble constant found by this technique might be attributed to
projection or selection effects, calibration uncertainties, unresolved
substructure, or some other problem, but it is clear that a
substantial fraction of the \textit{random} component of the error is
due to uncertainty in the X-ray temperature.  A deep X-ray observation
with \xmm, which can detect arriving cluster photons of energy up to
about 10~keV, and so can make a better measurement of the temperature
of a hot cluster such as \cl\ (which Hughes \& Birkinshaw 1998b found
to have $kT \approx 7.55 \ \rm keV$ in the rest frame), would be
expected to resolve this problem.

Thus we obtained a long \xmm\ observation of \cl\ with the aim of
improving the measurement of the temperature and metal abundance of
the cluster, and hence making it more useful as a cosmological
tracer. This paper describes the data, their analysis, and the
implications for the temperature and structure of \cl\ and one of the
companion clusters in the supercluster.  We also present the \xmm\
spectrum of the companion quasar.  We revise the estimate of the
Hubble constant based on \cl\ using our new temperature for the
cluster, and compare an estimate of the total gravitating mass with
that measured from gravitational lensing. We compare the gas mass
fraction in \cl\ with the value for the Universe as a whole given by
$\Omega_{\rm b}/\Omega_{\rm m}$.  Finally, we point out structural
features of \cl\ which require further study.

We use a cosmology in which $\Omega_{\rm m} = 0.3$ and
$\Omega_\Lambda = 0.7$, and we adopt
$H_o = 70$~km s$^{-1}$ Mpc$^{-1}$ outside the discussion of
the use of \cl\ as a tracer of the Hubble flow.

\section{Observations}
\label{sec:obs}

\xmm\ observed \cl\ on December 30th, 2000.  In this paper we
concentrate on data from the European Photon Imaging Camera (EPIC)
which incorporates one pn (Sr\"uder et al.~2001) and two MOS (MOS1 and
MOS2; Turner et al.~2001) CCD-array cameras.  The cameras were
operated in Full Frame mode with the medium optical blocking
filter. The data were provided to us in the form of two separate
observations, of maximum livetime (for the MOS cameras) roughly
31.3~ks and 5.5~ks, separated by 110 minutes.  Analysis presented here
uses the latest software available from the \xmm\ project at
http://xmm.vilspa.esa.es/, {\sc sas v5.3.3}.  Calibrated event files
were generated using the {\sc emchain\/} and {\sc epchain\/} scripts
(described in detail on the project WWW site) and, after merging the
data for each camera separately for the two observations, we have
selected for analysis the good events with patterns 0 to 12 from the
MOS data and patterns 0 to 4 from the pn data.

The light curves extracted from the full-field, pattern=0,  10--12~keV
events were examined for intervals
of high particle background, which we defined as
those with $> 0.1$ ct s$^{-1}$ for MOS1 and MOS2, and
$> 0.2$ ct s$^{-1}$ for the pn. After removing the high-background
intervals, the remaining exposure
times were roughly 35.9~ks, 36~ks, and 24.3~ks for MOS1, MOS2, and pn,
respectively.

\begin{figure}
\epsfxsize 8.4cm
\epsfbox{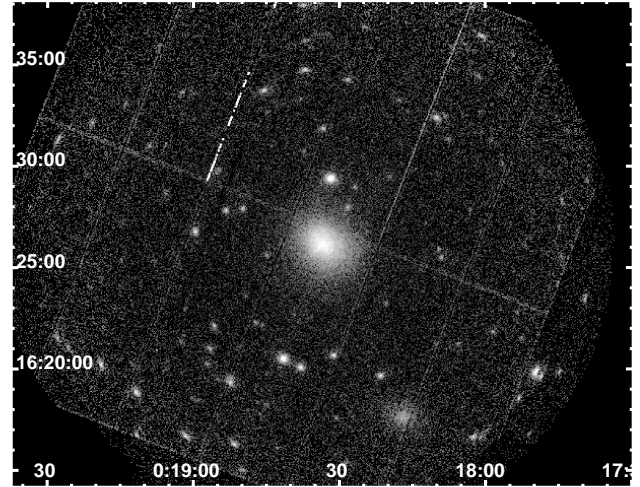}
\caption{
Combined MOS1, MOS2 and pn data for the observation of {\cl}~in the
0.3--5~keV energy band, with a logarithmic intensity scale, a pixel
size of 3~arcsec, and showing J2000 equatorial coordinates.  Chip
boundaries are visible. This image is vignetting corrected and without
background subtraction.  The bright source at a declination of roughly
$16^\circ 29' 26''$, north of \cl, is the quasar E~0015+162 (Margon et
al.~1983), and the poor companion cluster RX~J0018.3+1618 (Hughes 
et al.~1995) is visible to the SW.
} 
\label{largeimage}
\end{figure}

Our analysis follows the prescription in Arnaud et al.~(2002).  This
is superior to the simple use of local background in analysing
extended-source data because it treats separately the non-vignetted
particle background and the vignetted cosmic X-ray background.  First
the {\sc evigweight\/} task was run on both \cl\ and blank-sky data so
that after their subtraction the spectra and surface-brightness
profiles (with local background) that are extracted are corrected for
vignetting.  The vignetting-corrected MOS1, MOS2 and pn event files
have been combined for display purposes.  Figure~\ref{largeimage}
shows the 0.3--5 keV image of a wide field, centred on \cl, from this
merged event file.

We use the data compiled by Lumb (2002), taken from deep-sky exposures
cleaned of sources, to subtract from our data the particle background,
which is assumed to provide all the pattern=0 counts between 10 and 12
keV.  After applying our good-time-interval screening, the full-field,
pattern=0, 10--12~keV count rates are the same in the \cl\ and
blank-sky fields to within 1\%, 7\% and 16\%, for the MOS1, MOS2, and
pn, respectively.  The relevant scaling factor is applied to the
vignetting-corrected blank-sky data before subtraction from the image
data.

The blank-sky data are not expected to be a good representation of the
background in the \cl\ data at low energies. Here the cosmic X-ray
background is the dominant contributor and, more importantly, at
energies below about 1~keV the medium optical blocking filter used for
our observations has a lower transparency than the thin filter used
for the blank-sky data.  This results in an over-subtraction of
background at the lowest energies.  This is corrected for by using a
blank-sky-subtracted source-free region of the \cl\ image as a
background for the blank-sky-subtracted on-source data.  By selecting
this background region to be an annulus of radii 4 and 11 arcmin,
between position angles 90 and 120 degrees, and 220 and 310 degrees,
we limited the number of source-contaminated exclusion regions to one
circle of radius 30 arcsec.

As data have been corrected for vignetting, spectral fitting uses 
on-axis response matrices and effective areas.  We have used the
version 6 response matrices made available by the project,
and have generated on-axis effective areas using the {\sc arfgen}
task.  Bins between energies of 0.4 and 10 keV are
included in spectral fitting using the {\sc xspec} software.

\section{Emission-weighted Cluster Temperature}
\label{sec:temp}

For good signal to noise, we have found the cluster spectral
parameters using the net counts (roughly 32,600 counts between 0.4 and
10 keV) extracted from a circle of radius 1.5 arcmin.  We fitted the
data to a redshifted {\sc mekal} model modified by local Galactic
absorption. We let the relative normalization between the MOS1, MOS2
and pn data be free but all other parameters are in common.  This
results in MOS1 and MOS2 normalizations in good agreement, but the pn
normalization is a few per cent lower, consistent with the
evaluation of the EPIC flux calibration by Saxton (2002).

We obtain a value of $N_{\rm H} = (4.3 \pm 0.6) \times 10^{20}$
cm$^{-2}$ (95\% confidence for one interesting parameter) for the
Galactic hydrogen column density when it is a free parameter, in
agreement with the value of $4.04 \times 10^{20}$ cm$^{-2}$ found from
21~cm observations (Dickey \& Lockman 1990).  The value of the
redshift fitted from the X-ray spectrum, $z =
0.532^{+0.011}_{-0.013}$, is slightly low as compared with published
optical values of 0.5455 (Dressler \& Gunn 1992) and 0.5481 (Ellingson
et al.~1998), suggesting that, although the EPIC energy scale is
believed to be calibrated to better than 5~eV (Kirsch 2002), there may
be a small additional systematic error in the \cl\ observations.  We
have fixed the Galactic absorption to the 21~cm value but have allowed
the cluster redshift to be free in finding the cluster parameters and
uncertainties reported in this paper.

\begin{figure}
\epsfxsize 5.5cm
\epsfbox{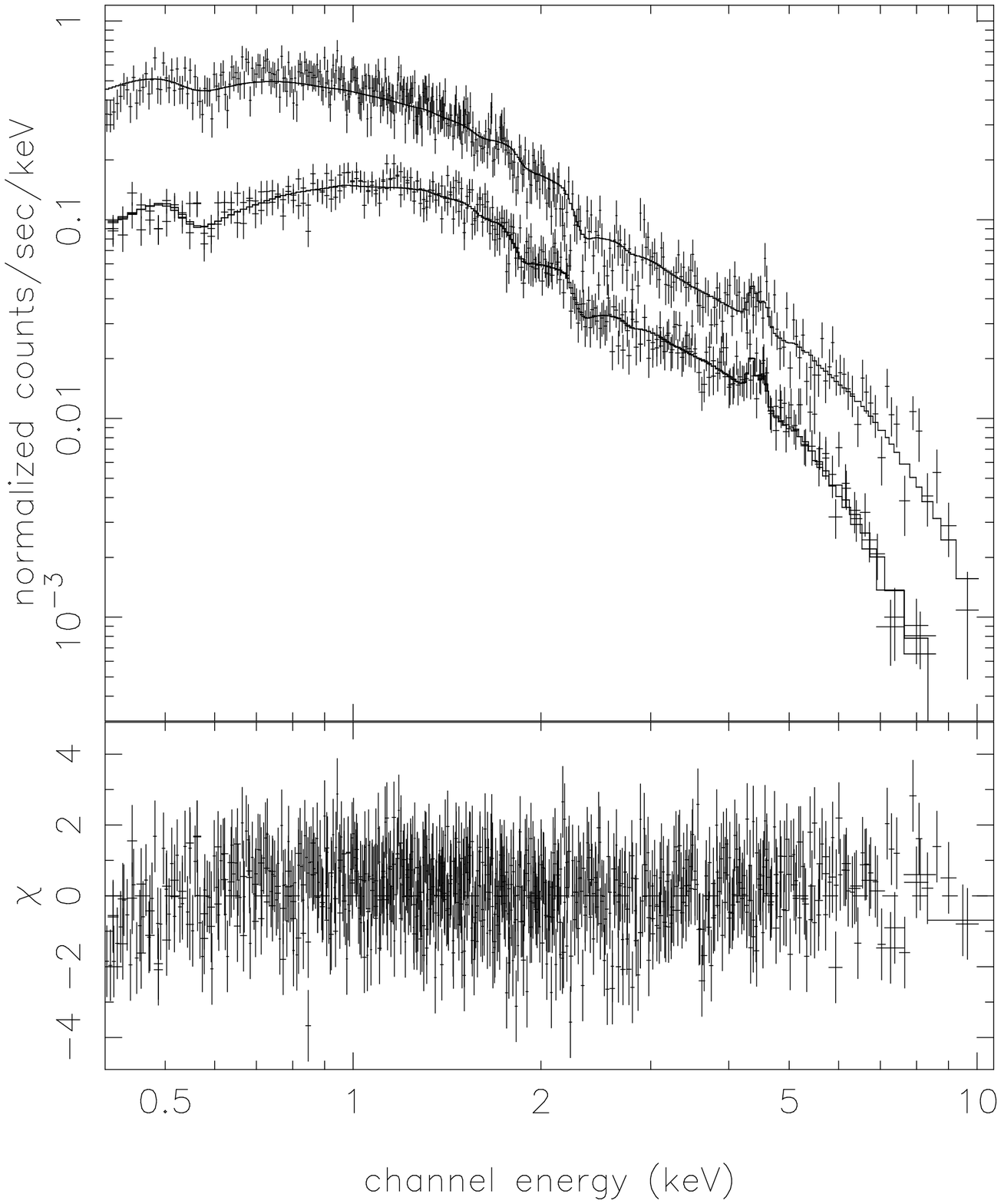}
\caption{
\xmm\ EPIC spectrum of \cl\ from counts within a circle of 
radius 1.5 arcmin.  The upper spectrum is from the pn, and the lower
spectrum is an overlay of MOS1 and MOS2 data points.
The fit is to an isothermal model with kT = 9.1~keV and an
abundance of 0.22 times the solar value.
} 
\label{spectrum}
\end{figure}

\begin{figure}
\epsfxsize 4.5cm
\epsfbox{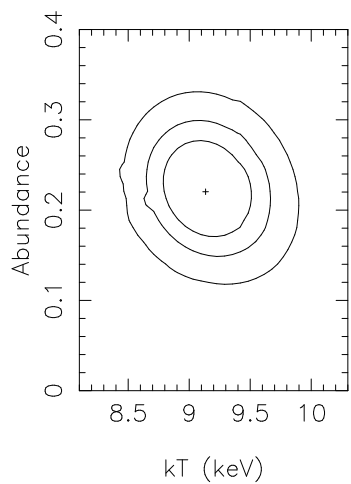}
\caption{
\cl\ spectral-parameter uncertainty contours using counts within a circle of 
radius 1.5 arcmin.  Contours are $1\sigma$, 90\% and 99\% for two
interesting parameters, with the normalizations and redshift
allowed to vary.
} 
\label{speconts}
\end{figure}

The best fit gives $kT = 9.13^{+0.24}_{-0.22} (^{+0.49}_{-0.44})$~keV
and an abundance of $0.22^{+0.04}_{-0.03} (^{+0.07}_{-0.065})$ times
the solar value, where uncertainties are 
statistical and quoted as $1\sigma$ (unbracketed), with 95\% errors given in brackets,
for one interesting parameter.  The model is shown together with the data in
Figure~\ref{spectrum}, and joint-confidence uncertainty contours are
shown in Figure~\ref{speconts}.  With a reduced $\chi^2$ of 1.09
($\chi^2 = 859$ for 786 degrees of freedom) the null hypothesis
probability is only 4\%, but this is likely to result from the high
statistical precision of the data coupled with small remaining
systematic calibration uncertainties (Saxton 2002).  An apparently
random scatter in the residuals to the pn data is diminished if the
pattern=0 events only are used for the pn, making the combined MOS and
pn fit acceptable (reduced $\chi^2$ of 1.01: $\chi^2 = 706$ for
697 degrees of freedom).  However, as the best-fit parameter values
are unchanged, we elect to include patterns 0--4 in the fits to
improve statistical accuracy.

There is consistency between the parameter values
fitted separately for data from the MOS
($kT = 9.3^{+1}_{-0.7}$, abundance $= 0.22^{+0.09}_{-0.08}$ times the
solar value, 
$z=0.534^{+0.015}_{-0.013}$, $N_{\rm H} = 4.4^{+0.9}_{-1.0} 
\times 10^{20}$ cm$^{-2}$)
and pn ($kT = 8.6^{+0.8}_{-0.6}$ , abundance $= 0.22^{+0.1}_{-0.09}$
times the solar value, 
$z=0.525^{+0.02}_{-0.03}$, $N_{\rm H} = 4.3^{+0.7}_{-0.8} 
\times 10^{20}$ cm$^{-2}$), 
all quoted with 95\% uncertainties for one interesting parameter.

The observed 2--10~keV flux within the circle of 1.5~arcmin radius
is measured to be
$1.43 \times 10^{-12}$ ergs cm$^{-2}$ s$^{-1}$ and
$1.39 \times 10^{-12}$ ergs cm$^{-2}$ s$^{-1}$
with the MOS and pn cameras, respectively.
Between 0.4 and 10 keV the values are
$2.24 \times 10^{-12}$ ergs cm$^{-2}$ s$^{-1}$ and
$2.18 \times 10^{-12}$ ergs cm$^{-2}$ s$^{-1}$, respectively.
The average volume-weighted emission measures
($\int n_{\rm e} n_{\rm p} dV/4 \pi D_{\rm L}^2$) are
$(1.22 \pm 0.02) \times 10^{11}$ cm$^{-5}$ and
$(1.18 \pm 0.02) \times 10^{11}$ cm$^{-5}$, respectively
($1\sigma$ uncertainties)\footnote{
In the {\sc xspec} X-ray
spectral-fitting software used here, and by most authors, the
normalization given by fitting thermal plasma models
is $10^{-14} \times (1+z)^2$ times the
volume-weighted emission measure.}.  

\section{Gas Distribution}
\label{sec:distrib}

\begin{figure}
\epsfxsize 8.0cm
\epsfbox{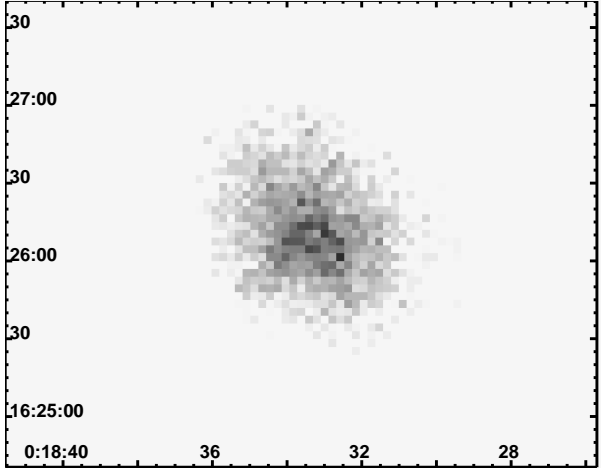}
\caption{
Same as Figure 1, showing only the centre of {\cl}. Scale is linear
with black corresponding to about 70 counts and the faintest pixels
containing about 20 counts. An inverted V-shaped filamentary structure
is apparent.  This is also seen in the individual MOS and pn images,
and is unrelated to chip orientation.
} 
\label{centreimage}
\end{figure}

The intensity distribution of X-ray emission in the central part of
\cl\ (Fig.~\ref{centreimage}) shows an apparent filamentary structure
in the shape of an inverted `V'.  A true colour image, made using
software from the \chandra\ project (the {\sc ciao dmimg2jpg}
task), with counts in 0.3--1~keV, 1--2~keV, and 2--8~keV corresponding
to red, green and blue, respectively, suggests that the emission at
the centre of the `V' may be hotter than in the surrounding part of
the cluster. However the prong to the SE appears cooler than average.

These spectral differences also appear when we use the {\sc
iraf/stsdas} {\sc ellipse} program to model the underlying large-scale
structure of the cluster, and then subtract a scaled version of the
model first from a 0.3--1.5~keV image and then a 2--8~keV image.  The
most prominent positive residuals in the soft image track the SE prong
of the inverted `V', whereas in the hard image they are at the cluster
centre.  The diffuse asymmetric emission to the west of the cluster,
at about 1.5~arcmin from the centre (Fig.~1), appears somewhat softer
than \cl\ as a whole.  This western diffuse emission was previously
discussed by Neumann \& B\"ohringer (1997) using \rosat\ data, and is
interpreted as a merging subcomponent of the cluster that is
associated with a peak in the weak-lensing mass map of Smail et
al.~(1995).

\begin{figure}
\epsfxsize 4.5cm
\epsfbox{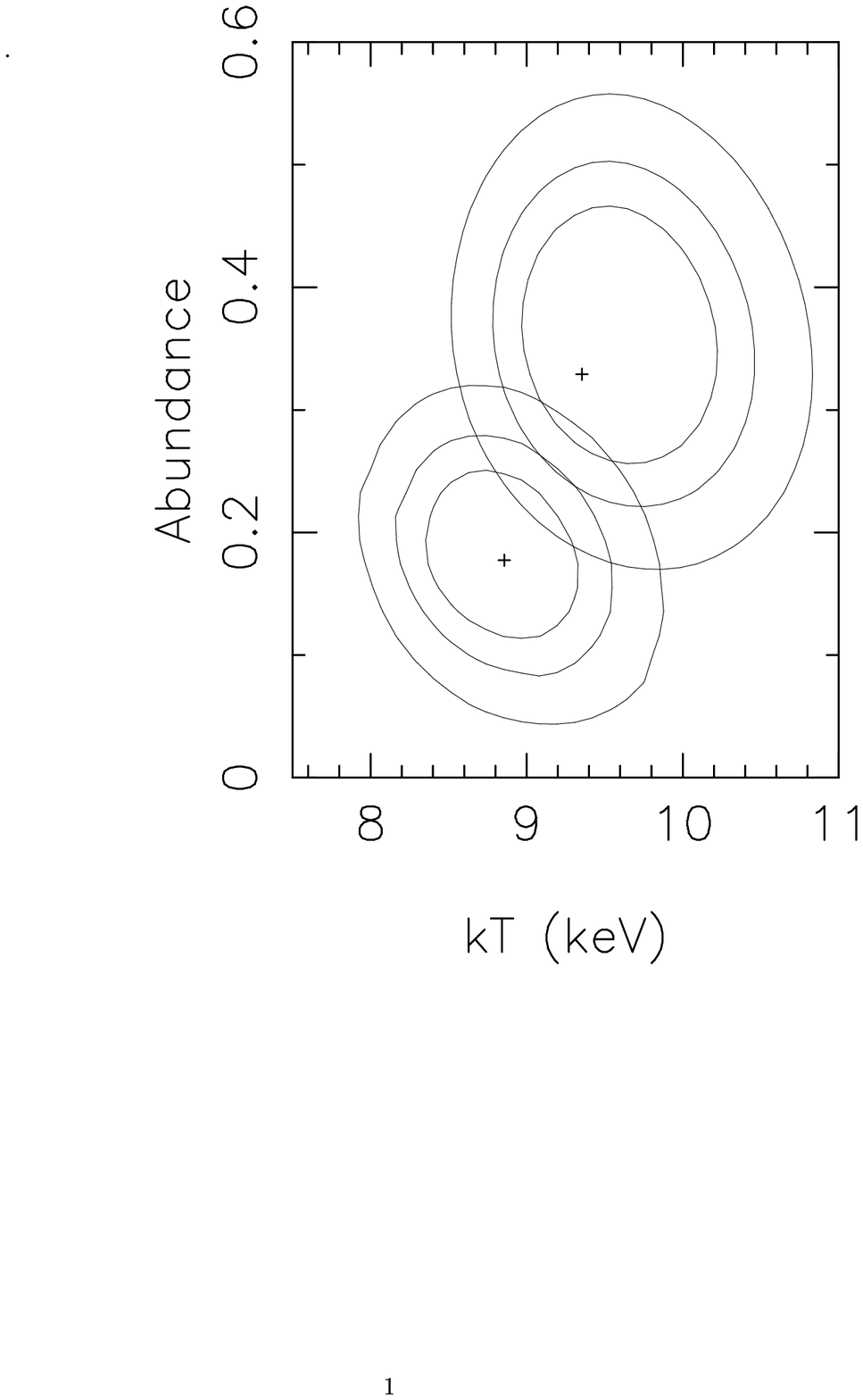}
\caption{
Same as Figure~3, separating counts within a circle of radius 0.6 arcmin
(upper right contours) from those within an annulus of radii
0.6 and 1.5 arcmin. 
} 
\label{specinoutconts}
\end{figure}

We have fitted separately the counts within a source-centred circle of
radius 0.6 arcmin and a source-centred annulus of radii 0.6 and 1.5
arcmin using the method and model described in \S\ref{sec:temp}.  The
results (Fig.~\ref{specinoutconts}) indeed show that the inner region
is hotter and of higher abundance than the outer one, at a combined
confidence of 99\% for two interesting parameters.  The evidence for
spatial and temperature structure in the inner regions may result from
merger activity.  A deeper observation would be required to probe this
in detail.

We have extracted the circularly-symmetric radial profile of
background-subtracted 0.3-5~keV counts and fitted it to a spherical,
isothermal, $\beta$ model of surface brightness $b_{\rm x}(\theta)
\propto (1 + [\theta/\theta_{\rm c}]^2)^{-3\beta + 0.5}$, convolved
with the point spread function (PSF).  While the cluster is clearly
ellipsoidal in appearance, and displays hotter regions within its
core, the value of $\beta$ and the average core radius of the model
produced by fits of spherical functions are close to those obtained by
making more complicated ellipsoidal fits (see Tables~1 and~2 in Hughes
\& Birkinshaw 1998b). If the fit results are to be used in the
calculation of the distance of the cluster (as, for example, in
\S\ref{sec:h0}), then values obtained from ellipsoidal fits could not
be used directly to estimate $H_0$, since the projection is unknown
for any individual cluster. However, $H_0$ estimates obtained from
spherical model fits for a {\it population} of clusters are almost
unbiased, provided that the clusters are selected in an
orientation-unbiased fashion. Thus spherical modeling of clusters is
sufficient for the purpose of building up a sample of distance
measures from clusters.

To construct the radial profile we have summed the counts from the
three cameras.  For the PSF we have used the on-axis parameter values
for 1~keV from Ghizzardi's (2001, and private communication in 2002)
fits of a $\beta$ model to data from stellar sources.  We have summed
the MOS1, MOS2, and pn PSFs, weighted by each camera's contribution to
the counts from \cl.  The counts from a circle of radius 20 arcsec
around the quasar 3.25 arcmin to the north of \cl\
(Fig.~\ref{largeimage}) are excluded from the fits.

\begin{figure}
\epsfxsize 8.5cm
\epsfbox{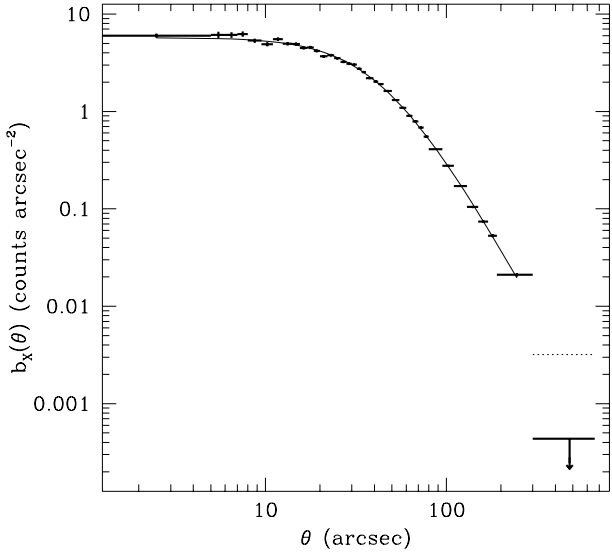}
\caption{
0.3--5~keV radial profile of {\cl} after subtraction of deep-sky and
local backgrounds, as described in \S\ref{sec:obs}.  The curve is
the best fit circularly symmetric $\beta$ model convolved with the
PSF: $\beta = 0.697$, $\theta_{\rm c} = 36.546$~arcsec.
The horizontal dotted line shows the contributions, taken into account
in the fitting, of the model to the local background annulus. 
} 
\label{profile}
\end{figure}

The radial profile and best-fit model are shown in Figure~\ref{profile}.
While the filamentary structure seen in Figure~\ref{centreimage} leads
to a poor fit within about 20 arcsec of the centre, this has
negligible effect on the values found for $\beta$ and $\theta_{\rm
c}$, and the overall fit is acceptable ($\chi^2 = 50.3$ for 37 degrees
of freedom).  The values found are $\beta = 0.697 \pm 0.010$, and
$\theta_{\rm c} = 36.6 \pm 1.1$ arcsec, where, since the parameters
are strongly correlated, the errors quoted are one sigma for two
interesting parameters.  Nothing in the radial profile or spectral
properties of \cl\ is suggestive of a cooling flow.

\section{Companion Cluster, RX~J0018.3+1618}
\label{sec:companion}

\begin{figure}
\epsfxsize 5.5cm
\epsfbox{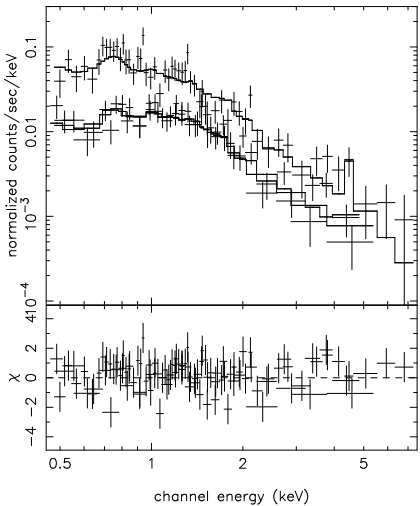}
\caption{
\xmm\ EPIC spectrum of the companion cluster from counts within a circle of 
radius 1.5 arcmin.  The upper spectrum is from the pn, and the lower
spectrum is an overlay of MOS1 and MOS2 data points.
The fit is to an isothermal model with kT = 3.8~keV and an
abundance of 0.6 times the solar value.
} 
\label{spectrumcomp}
\end{figure}

The companion cluster to the SW of \cl\ (Fig.~\ref{largeimage})
appears redder than \cl\ in a true colour image (made as in
\S\ref{sec:distrib}).  We have found spectral-parameter values using
on-source net counts from a circle of radius 1.5 arcmin (roughly 3,200
counts between 0.4 and 10 keV) in a similar manner to the extraction
and fitting for \cl\ itself.  The fit to a {\sc mekal} model is good,
with $\chi^2 = 109$ for 101 degrees of freedom.  The column density of
$N_{\rm H} = 5.8^{+2.6}_{-2.3} \times 10^{20}$ cm$^{-2}$ ($1\sigma$
errors) is consistent with the value from 21~cm measurements ($4.09
\times 10^{20}$ cm$^{-2}$; Dickey \& Lockman 1990), and so we fix the
value to the radio-derived measurement.  Figure~\ref{spectrumcomp}
shows the data and model whose best-fit parameter values and
uncertainties are as follows, where the latter are $1\sigma$ (95\%)
confidence for one interesting parameter where all the other parameter
values are allowed to vary: $kT = 3.8^{+0.3}_{-0.3}(^{+0.8}_{-0.6})$,
abundance $= 0.6^{+0.3}_{-0.2}(^{+0.6}_{-0.4})$ times solar,
$z=0.472^{+0.062}_{-0.022}(^{+0.09}_{-0.05})$.  The redshift is
consistent (at 95\% confidence, or $1\sigma$ if the systematic error
of 0.016 suggested for \cl\ is applied) with the value of $0.5506\pm
0.0012$ found from optical measurements by Hughes et al.~(1995) who
were able only to place a lower limit on the cluster temperature ($kT
> 1.5$ keV in the observed frame) using data from ROSAT.  The
precision with which the spectral parameters for the companion cluster
are measured is similar to that with which the parameters for \cl\
were known prior to this work (Furuzawa et al.~1998; Hughes \&
Birkinshaw 1998b).

The net observed 2--10~keV and 0.4--10 keV fluxes are measured to be 
$7.7 \times 10^{-14}$ ergs cm$^{-2}$ s$^{-1}$ and
$1.6 \times 10^{-13}$ ergs cm$^{-2}$ s$^{-1}$, respectively.
The volume-weighted emission measure is 
$(1.09\pm 0.12) \times 10^{10}$ cm$^{-5}$.

Because this cluster lies on a gap between CCDs in the pn camera, as
can be seen in Figure~\ref{largeimage}, we have fitted the radial
profile using only data from the MOS cameras.  The parameter values
found for a $\beta$ model, convolved with the PSF corresponding to the
off-axis position of the cluster, are of comparable value and
precision to the values found by Hughes et al.~(1995) using \rosat\
{\sc PSPC} data.  We find $\beta = 0.74^{+0.36}_{-0.17}$, $\theta_{\rm
c}= 28^{+15}_{-9}$ arcsec ($1\sigma$ uncertainties).

\section{Companion Quasar, E~0015+162}
\label{sec:agn}
\begin{figure}
\epsfxsize 5.5cm
\epsfbox{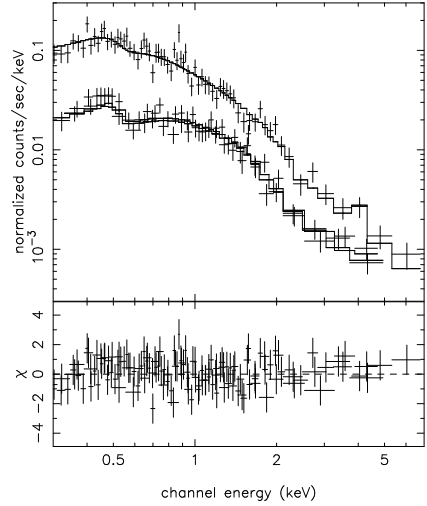}
\caption{
\xmm\ EPIC spectrum of the companion quasar from counts within a circle of 
radius 40 arcsec.  The upper spectrum is from the pn, and the lower
spectrum is an overlay of MOS1 and MOS2 data points.
The fit is to a power law of photon spectral index 2.38,
with a narrow Fe fluorescence line of rest energy 6.4 keV.
} 
\label{spectrumagn}
\end{figure}

The companion radio-quiet quasar of redshift 0.554 (Margon et
al.~1983) to the north of \cl\ (Fig.~\ref{largeimage}) has a
relatively steep spectrum, and we use the net counts between 0.3 and
10 keV within a source-centred circle of radius 40 arcsec for our
fits.  A fit to an absorbed power law is acceptable ($\chi^2 = 113$
for 126 degrees of freedom), with the column density consistent with
the Galactic value from 21~cm measurements.  $\chi^2$ is reduced by
7.1 for one extra parameter (significant at 99\% confidence on an F
test) when a narrow Fe fluorescence line, modeled as a Gaussian of
fixed rest-frame energy 6.4~keV and $\sigma = 10$~eV (small compared
with the roughly 130~eV FWHM resolution of EPIC at the observed
energy) is included in the model.  If the line energy is allowed to be
a free parameter, the best-fit value and $1\sigma$ error is
$6.1^{+0.6}_{-0.1}$.  Fixing the line energy at 6.4~keV, the following
parameter values and uncertainties are found, where the latter are
$1\sigma$ (95\%) confidence for one interesting parameter where all
the other parameter values are allowed to vary: power-law photon index
$\Gamma = 2.38 \pm 0.06 (\pm 0.13)$, $N_{\rm H} = 3.4 \pm 1.0 (\pm
1.9) \times 10^{20}$ cm$^{-2}$, equivalent width of the Fe line $=
308^{+203}_{-123} (^{+352}_{-271})$ eV.  The spectrum is shown in
Figure~\ref{spectrumagn}.  The observed 0.5--10 keV flux from the
quasar is $1.61 \pm 0.07 \times 10^{-13}$ ergs cm$^{-2}$ s$^{-1}$
($1\sigma$ error) and the luminosity is $2.15 \times 10^{44}$ ergs
s$^{-1}$.

The AGN is roughly 8\% the strength of \cl\ (in a cluster-centred
circle of radius 4~arcmin) in terms of net counts between 0.4 and 10
keV (roughly 3,500 counts from the AGN).  Reeves et al.~(1997) fitted
the \asca\ spectrum of the AGN and found only an upper limit for
narrow Fe fluorescence-line emission, but the \asca\ spectrum would
have been heavily contaminated with emission from \cl, and this would
seem to account for the inconsistency between our results and the flat
spectral index ($1.99 \pm 0.09$) and large flux ($2.92 \times
10^{-12}$ ergs cm$^{-2}$ s$^{-1}$ between 0.5 and 10 keV) found by
these authors.  Our spectral index is consistent with the value of
$2.54 \pm 0.18$ ($1\sigma$ error) found for the quasar by Hughes \&
Birkinshaw (1998b) from a combined thermal and power-law model fitted
jointly to \asca\ and \rosat\ PSPC data in order to characterize the
emission both from \cl\ and the quasar.  Their power-law normalization
is also consistent with our better-determined value.  \xmm\ has
improved the precision with which the spectral index is measured by a
factor of three, and has provided the first evidence for an Fe
fluorescence line in this radio-quiet quasar.

\section{Hubble constant and Gas Content of \cl}
\label{sec:h0}

\xmm\ has measured the temperature of \cl's atmosphere to $\pm 2.5\%$.
This is a considerable improvement on the $9\%$ uncertainty
obtained using \asca\ and \rosat\ (Hughes \& Birkinshaw 1998b), and so
reduces one of the main sources 
of random error in the Hubble constant as deduced by comparing the
X-ray and Sunyaev-Zel'dovich effect properties of the cluster.

The \xmm-derived emission measure within 1.5~arcmin of the
centre, $(1.20 \pm 0.03) \times 10^{11} \ \rm cm^{-5}$, from the
weighted average of the MOS and pn values, implies an
integrated cluster emission measure of 

\begin{equation}
 { \int n_{\rm e} \, n_{\rm p} \, dV  \over 4 \pi D_{\rm L}^2 }
 =
 (1.75 \pm 0.05) \times 10^{11} \ \rm cm^{-5}
 \label{eq:emvalue}
\end{equation}
if the fraction of the emission contained within a 1.5-arcmin radius
cylinder is extrapolated to the full cluster using the best-fit radial
profile parameters. This emission measure is related to the angular
diameter distance of the cluster, $D_{\rm A}$, and the central
electron density, $n_{\rm e0}$, by 

\begin{equation}
 {\rm EM} = {D_{\rm A} \, \theta_{\rm c}^3 \, n_{\rm e0}^2 \over
            \eta \, (1 + z)^4 } \, {\sqrt{\pi} \over 4} \,
           {\Gamma\left(3\beta-{3 \over 2}\right) \over
            \Gamma\left(3\beta\right)}
            \quad
 \label{eq:eminfinity}
\end{equation}
if the \xmm-derived $\beta$ model is used to describe the overall
structure of the cluster, and so the X-ray data provide a measurement
of $D_{\rm A} n_{\rm e0}^2$. The symbol $\eta$ in equation (2) is the
electron/proton ratio, $\approx 1.17$.  The emission measure
(\ref{eq:emvalue})
is less than half the value given by Hughes \& Birkinshaw (1998b),
because of an error in the way that plasma thermal emission
normalizations were reported by {\sc XSPEC}.
Since the emission measure was not used in the
calculation of the cluster distance in the Hughes \& Birkinshaw paper,
however, the conclusions about the cluster distance there are
unchanged by this error.

The $\beta$ model can also be used to calculate the central
Sunyaev-Zel'dovich effect for the cluster,

\begin{eqnarray}
 \Delta T_0 &=& - 2 \, T_{\rm rad} \, \sigma_{\rm T} \, 
                    \left( {k T
                         \over m_{\rm e} c^2} \right) \, 
                    n_{\rm e0} \, D_{\rm A} \, \theta_{\rm c}
                    \sqrt{\pi} \, {\Gamma\left({3 \over 2}\beta -
                    {1\over 2}\right) \over 
                    \Gamma\left({3 \over 2}\beta\right)} \quad
 \label{eq:dt0}
\end{eqnarray}
where $T_{\rm rad}$ is the thermodynamic temperature of the microwave
background radiation and $\sigma_{\rm T}$ is the Thompson scattering
cross-section. 

The central Sunyaev-Zel'dovich effect of \cl\ is found by fitting a
one-dimensional profile (Hughes \& Birkinshaw 1998b) or a
two-dimensional map (Reese et al. 2000; Grainge et al. 2002) of the
effect. Using slightly different $\beta$ models for the cluster
atmosphere, values $\Delta T_0 = -1.20 \pm 0.19$, $-1.24 \pm 0.11$,
and $-1.08 \pm 0.11 \ \rm mK$ were found in these investigations.
We reanalysed the Hughes \& Birkinshaw (1998b) data using the
$\beta$-model parameters based on our \xmm\ data, and scaled the
interferometric data (which are relatively insensitive to the
$\beta$-model parameters along degeneracy lines in the [$\beta$,
$\theta_{\rm c}$] plane) to these same parameters. A weighted
combination of all three independent values leads to a central
Sunyaev-Zel'dovich effect for \cl\ of $\Delta T_0 = -1.26 \pm
0.07 \ \rm mK$.  Via equation~(\ref{eq:dt0}), this implies a
measurement of $D_{\rm A} n_{\rm e0}$.

Combining our measurements of $D_{\rm A} n_{\rm e0}^2$ and $D_{\rm A}
n_{\rm e0}$, and using the optical cluster redshift of 0.5481
(Ellingson et al.~1998), we find the angular diameter distance and
central electron density of \cl\ to be
\begin{eqnarray}
 D_{\rm  A} &=&  1.36 \pm 0.15                 \ \rm Gpc, \ and
                 \label{eq:da} \\
 n_{\rm e0} &=& (8.8  \pm 0.5 ) \times 10^{-3} \ \rm cm^{-3} 
                 \quad . \label{eq:ne} 
\end{eqnarray}
In a $\Lambda$CDM cosmology with $\Omega_{\rm m} = 0.3$,
$\Omega_{\Lambda} = 0.7$, this corresponds to a Hubble constant of $68
\pm 8 \ \rm km \, s^{-1} \, Mpc^{-1}$ (random error only), which is
about $45\%$ larger than the values estimated by Hughes \& Birkinshaw
(1998b), Reese et al.~(2000), and Grainge et al.~(2002). The change in
$H_0$ arises about equally from the increased best-fit gas temperature
found in the present \xmm\ study and the change from a flat
matter-dominated cosmology to a flat $\Lambda$CDM cosmology.

In addition to the statistical error in equation~(\ref{eq:da}), 
the distance is
subject to systematic errors, the largest of which are 5\%
calibration uncertainties for the Sunyaev-Zel'dovich effect and
X-ray normalization, a 20\% uncertainty from
possible projection effects, and a 15\% uncertainty from substructure
within the cluster (Hughes \& Birkinshaw 1998b). Overall, therefore, we
deduce an angular diameter distance for \cl\ of $1.36 \pm 0.15({\rm
statistical}) \pm 0.34({\rm systematic}) \ \rm Gpc$, which implies a
Hubble constant of $68 \pm 8({\rm statistical}) \pm 18({\rm
systematic}) \ \rm km \, s^{-1} \, Mpc^{-1}$. This is consistent with
the value of the Hubble constant found in the HST distance-scale
project (Mould et al. 2000), but the systematic errors are too large
for a useful independent measurement of $\Omega_\Lambda$ to result
from our distance for \cl.

It is interesting to use the X-ray data for \cl\ to estimate
the mass of the cluster within the region well-probed by \xmm. For an
isothermal gas in hydrostatic equilibrium in a spherical cluster, the
total gravitating mass within a cylinder of radius $r$ about the
cluster centre is 
\begin{equation}
 M_{\rm tot}(r) = {3 \pi \beta \over 2} \, {k T
 \over \mu \, m_{\rm p} \, G} \, {r^2 \over r_{\rm c}} \, \left( 1 +
 {r^2 \over r_{\rm c}^2} \right)^{-{1 \over 2}} \quad .
\end{equation}
For $r = 248 \ \rm kpc$, this corresponds to a mass of $(2.0 \pm 0.1)
\times 10^{14} \ \rm M_\odot$. This value is consistent with the mass
of $(2.7 \pm 0.9) \times 10^{14} \ \rm M_\odot$ measured from the
gravitational lensing effect of the cluster within the same
cylindrical radius (Smail et al. 1997; value converted to our adopted
cosmology).

We can compare this total gravitating mass with the gas mass within
the same cylindrical radius, based on the central electron density
that we measure. The result, $M_{\rm gas}({248 \ \rm kpc}) = (2.6 \pm
0.2) \times 10^{13} \ \rm M_\odot$ indicates that X-ray emitting
gas contributes a fraction $(0.13 \pm 0.02)$ of the total mass of
the cluster in this 248-kpc radius cylinder (for $H_0 = 70 \ \rm km \,
s^{-1} \, Mpc^{-1}$), or $(0.13 \pm 0.04)$ \textit{independent of
cosmology} if we use the cluster distance in equation~(\ref{eq:da})
and the electron density in equation~(\ref{eq:ne}).

The measured gas mass fraction is consistent with the value
$0.14^{+0.09}_{-0.04}$ within a radius of 65 arcsec (415 kpc) which is
obtained by correcting the result of Grego et al. (2001) to the
cosmology that we use and cluster temperature that we measure.  Our
temperature makes \cl's gas mass fraction more consistent
with the average in the Grego et al. sample, and with the averages for
the Myers et al. (1997) and Mohr et al. (1999) samples. Since these
samples derived the gas mass fraction using different analyses of the
available X-ray and/or Sunyaev-Zel'dovich effect data, their agreement
about the gas mass fraction and its lack of change with redshift
suggests that the assumptions used, notably the lack of small-scale
density or thermal structure in the atmospheres, are not seriously in
error. Since the baryonic matter content of the X-ray gas dominates
that of galaxies in clusters, and $\Omega_{\rm b}/\Omega_{\rm m} =
0.12 \pm 0.02$ (Turner 2002) for the Universe as a whole, it appears
that our 248-kpc radius cylinder through \cl, and clusters of galaxies
in general, are fair samples of the matter content of the
Universe. Studies of how the gas mass fraction in clusters varies with
redshift are therefore unlikely to provide useful information on
cluster evolution, at least at the current level of accuracy of the
measurements.

\section{Summary}
\label{sec:summary}

With the EPIC camera of \xmm\ we have measured the temperature and
distribution of the intra-cluster medium of \cl\ to unprecedented
precision.  The emission-weighted temperature within 1.5 arcmin of the
cluster centre is $kT = 9.13^{+0.24}_{-0.22}$~keV and the abundance is
$0.22^{+0.04}_{-0.03}$ ($1\sigma$ uncertainties).  The higher best-fit
temperature than found in previous work using \asca\ (Furuzawa et
al.~1998; Hughes \& Birkinshaw 1998b) affects the cluster's position
on the temperature-luminosity plane, now moving it into good agreement
with the $L_{\rm x} - T$ relation for nearby clusters (see figure~7 of
Schindler 1999).

Although there is some ellipticity in the plane of the sky, as
previously reported based on \rosat\ data (Neumann \& B\"ohringer
1997; Hughes \& Birkinshaw 1998), we find an acceptable fit of the
overall envelope out to 5 arcmin (1.9~Mpc) to a circularly-symmetric
$\beta$ model, representative of gas in hydrostatic equilibrium, with
$\beta = 0.697 \pm 0.010$, and core radius $\theta_{\rm c} = 36.6 \pm
1.1$ arcsec (where errors are $1\sigma$ for two interesting
parameters since $\beta$ and $\theta_{\rm c}$ are strongly
correlated).

While there is no spectral or spatial evidence to suggest a cooling
flow in \cl, an asymmetric enhancement to the west of the cluster is
apparent in the \xmm\ image.  Such an asymmetry was previously
reported by Neumann \& B\"ohringer (1997) and interpreted as a merging
subcomponent of the cluster.  The \xmm\ data also show a more central
asymmetric X-ray structure which may have a harder spectrum than the
cluster as a whole, and be evidence of merger activity.  A deeper
X-ray exposure would be required to probe this in more detail.

The improved spectral and spatial data provided by \xmm\
have allowed us to reduce the random error on the
angular diameter distance of \cl\ to 11\%. The resulting
estimate of the value of the Hubble constant, $68 \pm 8 {\rm
(random)} \pm 18 {\rm (systematic)} \ \rm km \, s^{-1} \, Mpc^{-1}$,
is in good agreement with the value $71 \pm 6 \ \rm km \, s^{-1} \,
Mpc^{-1}$ from the HST distance-scale project (Mould et
al. 2001). The largest remaining systematic component of the error
in our estimate of the Hubble constant arises from projection and
substructure effects, and could be reduced by a factor $\sim 3$ by
applying the technique to a sample of $10 - 20$ clusters with
equally-good X-ray and Sunyaev-Zel'dovich effect data as \cl.

The total gravitating mass within a radius of 248~kpc of the cluster
centre is in good agreement with that found from gravitational lensing
over the same region (Smail et al.~1997), which supports the
assumption that the gas in \cl\ can be treated as isothermal and in
hydrostatic equilibrium.  The measured gas mass fraction of $0.13 \pm
0.02$ is in remarkable agreement with that given by the cosmological
parameters for the Universe as a whole, suggesting that \cl\ is a fair
sample of the matter content of the Universe.

The cluster \cl\ is believed to be the dominant member of a
supercluster at $z \sim 0.5$.  The nearest companion cluster and
relatively bright companion quasar E~0015+162 (which is of an X-ray
luminosity more typical of a local Seyfert galaxy) are both well
measured in the \xmm\ EPIC data.  In previous \asca\ data the emission
from the quasar is confused with that from \cl\, and so values
reported for the quasar based on that observation are unreliable.  We
find that the spectrum of E~0015+162 fits well a power law absorbed
only by the column density of gas through our Galaxy, with a
relatively steep power-law photon index of $\Gamma = 2.38 \pm 0.06$.
There is evidence for the detection of an Fe fluorescence line.

\xmm\ finds that the spectrum of the nearest companion cluster,
RX~J0018.3+1618, fits a thermal model with $kT =
3.8^{+0.3}_{-0.3}$~keV and abundance $= 0.6^{+0.3}_{-0.2}$ times solar,
These uncertainties for RX~J0018.3+1618 are comparable
to those of earlier measurements of \cl.
Deeper and broader X-ray observations of the
supercluster may provide important information on a massive structure
at an early phase of its evolution.

\section*{Acknowledgements}
This work is based on observations obtained with \xmm, an ESA science
mission with instruments and contributions directly funded by ESA
Member States and the USA (NASA).  The observation was made as part of
the \xmm\ Survey Science Centre (SSC) Guaranteed-Time programme.  We
thank Mike Watson for organizing the SSC and its scientific programme.

\label{lastpage}

\end{document}